# Short note on the Sirenia disappearance from the Euro-North African realm during the Cenozoic: a link between climate and Supernovae?

Cabral, F.[1], Cachão, M.[1], Jorge Agostinho, R.[1], Prista, G.[1]

[1] University of Lisbon, Campo Grande 1749-016, Lisboa, Portugal

Keywords: Sirenia, Supernovae, Cenozoic, Climate

**Abstract**
Sirenia are marine mammals that colonized the European shores up to 2.7 Ma. Their biodiversity evolution follows the climate evolution of the Cenozoic. However, several climate events, as well as the global climate trend of this Era are still struggling to be understood. When considering only Earth processes, the climate evolution of the Cenozoic is hard to understand. If the galactic environment is taken into account, some of these climate events, as well as the global climate trend, became more easily understood. The Milky Way, through Supernovae, may bring some answers to why Cenozoic climate had this evolution. With the assumption that SN can induced changes in Earth climate in long time scales, Sirenia disappearance from Europe would be a side effect of this process.

**Introduction**
Sirenia, an Order (Sirenia Illiger, 1811) of placental mammals that, along with cetaceans, represent the only mammals that evolved for a fully aquatic life (Clementz et al., 2009), and have the particularity of being the only herbivorous marine mammals (Domning, 2002), commonly named Seacow. They feed mainly on seagrass, although there are some differences between the two extant families, the trichechids and the dugongids. The Trichechidae family is not highly specialized when it comes to food and habitat. They live in fresh and sea-water, feeding on more than 60 species of marine plants, and can be considered, as suggested by Anderson (2002), opportunistic feeders on seagrass. The Dugongidae family is more specialized, feeding mainly on seagrass and is almost exclusively found in salt water. These characteristics are not exclusive of the four actual species, as shown by MacFadden et al. (2004), and can be traced back to the extinct species of both families.
They first appeared after the Early Eocene Climatic Optimum (53–49 Ma (Höntzsch et al., 2011)), with the oldest record of *Pezosiren portelli* from Jamaica (Domning, 2001a). The coastal shores of Europe and North Africa have records dating from the Lutetian (Caria, 1957; Crusafont-Pairó, 1973; Savage, 1976; Zalmout and Gingerich, 2012), showing that the Tethys Sea was colonized by these marine mammals in both its north and south margins.
Sirenians were abundant in Tethyan coastal waters, and also in the Paratethys and the Mediterranean, which evolved from it. The Sirenia that inhabited the European/North African shores are almost exclusively of the Dugongidae family (more than 95% of the European and North African fossil record belong to dugongids), meaning that they fed mainly on seagrass. Seagrasses are marine phanerogams that colonized the coastal waters around 100 Ma ago (Hemminga and Duarte, 2000). They show few changes in their evolutionary process and many of the genera found today can be traced back to the Eocene (like *Cymodocea nodosa* and *Posidonia* sp. from the Eocene of Paris) (Domning, 2001b; Bianucci et al., 2008).

After the Miocene Climatic Optimum (MCO) around 14 Ma ago (Böhme, 2003; Domingo et al., 2012), European and North African Sirenia declined in biodiversity (Prista et al., 2013; 2014a). Only one genus is known from the Late Miocene and Pliocene, the *Metaxytherium*, and only three species have been classified for this time interval, *M. medium*, *M. serresii* and *M. subapenninum*.

The Tortonian is the last age with Sirenia fossil records in the Northeast Atlantic shores, north of Portugal. In the Faluns Sea, northeastern France, the record is abundant until the Tortonian (Cottreau, 1928; Lécuyer et al., 1996).

For the Messinian age, Portugal appears to be the only place with a fossil record on this side of the Atlantic, with two records from Santa Margarida do Sado (Ferreira do Alentejo, Beja) and Vale de Zebro (Alvalade, Setúbal) (Estevens, 2000). The Mediterranean has a slightly richer fossil record, with fossils from Libya (Zalmout and Gingerich, 2012) and probably Italy (Moncharmont Zei and Moncharmont, 1987) and Spain (Sendra Saez et al., 1998). This scenario is not totally understood, because it was during the Messinian that the Messinian Salinity Crises (MSC) occurred, a period of closure of the Mediterranean and great ecological disturbance (Prista et al., 2013).

The Pliocene shows an increase in the fossil record for the Mediterranean Sea, and only one occurrence in the Atlantic, belonging to the Gulf of Cadiz on the Morocco shore (Ennouchi, 1954).

The Zanclean age has records of *M. serresii*, which first appeared during the Late Miocene of Libya (Zalmout and Gingerich, 2012), from Montpellier, France (Domning and Thomas, 1987; Pilleri, 1987), Sahabi Formation, Libya (Domning and Thomas, 1987) and Alicante, Spain (Sendra et al., 1999; Bianucci et al., 2008). During this age appeared the last European/North African sirenian species, the *M. subapenninum*, with the oldest records from the Early Zanclean of Italy (Sorbi and Vaiani, 2007; Tinelli et al., 2012) and the last known records dated to the Early Piacenzian (Bianucci et al., 2008). *M. serresii* ceases its appearance in the fossil record during the Zanclean age.

This decline in biodiversity, which culminated with Sirenia disappearance from European chores, has been explained by Prista et al. (2013; 2014a), and is strongly related to climate changes. In fact, the taxonomic palaeobiodiversity of the European and North African sirenians shows a close relation with the palaeoclimatic evolution of the Cenozoic (Prista et al., 2014a). The authors also found that Sirenia biodiversity seems to be strongly affected by variations in seasonality, coupled with other geological processes, such as sea level changes, and responded rapidly during the Oligocene, the Early-Middle Miocene and the Late Miocene-Pliocene (Prista et al., 2014a).

The climate system is a unique and open system with a multitude of interacting components and processes (Müller, 2010). However one question remains: which forces drive optimum climates and ice ages? Although several mechanisms must play a role in this Earth climate process, the galactic environment should not be neglected. During the last decade of the 20th century evidences liking galactic cosmic rays (GCR) to cloud cover were found through satellite observations (Marsh and Svensmark 2000b). This opened a great scientific debate, and today this direct relation is practically discarded. However that doesn't mean that GCR don't play a role in the Earth's climate. Their influence on the atmospheric electric current has been demonstrated by several studies (e.g. Tinsley and Yu, 2004; Serrano *et al.*, 2006; Tinsley, 2012), and the CLOUD project at CERN has been conducting experiences that have lead us to better understand this process (Kirkby, 2008; Duplissy *et al.*,

2010; Kirkby *et al*., 2012). Could the Milky Way affected Sirenia evolution?

**Cenozoic climate over the past 65 Ma**
During the Palaeogene several optimum climates took place: the Palaeocene-Eocene Thermal Maximum (PETM) (Sexton et al., 2011), the Early Eocene Optimum Climate (EEOC) (Höntzsch et al., 2011) and the Mid-Eocene Optimum Climate (MEOC) (Witkowski et al., 2012). These are all periods of less temperature gradient between high latitude regions and the tropics. During the Oligocene, the last epoch of the Palaeogene, the Antarctic continental glaciations (ACG) started (Zachos *et al*., 2001; Pollard and DeConto, 2005; Zachos *et al*., 2008; Cotton and Pearson, 2011; Diester-Haass *et al*., 2011). Although today we attribute the ACG to the strong influence of the Antarctic Circumpolar Current (ACC), when the ACG started, (around 34 Ma), the ACC was not formed yet (Lawver and Gahagan, 2003; Pfuhl and McCave, 2005). This means that a mechanism, other then the ACC, should have forced the beginning of the glaciations. The ACG were accompanied by a major global cooling. Oceans global temperature dropped ~2.5ºC (Héran et al., 2010); in Central Asia and Europe is observed a drop in temperatures and an increase in aridity (Abels et al., 2011; Costa et al., 2011); the Earth temperature gradient increased. After the Oligocene, started the Neogene period. It's first epoch, the Miocene, was warmer then the Oligocene, and the Earth temperature gradient got smaller again. During the Early and Middle Miocene an optimum climate took place, the Miocene Climatic Optimum (MCO) (Böhme, 2003; Kroh, 2007; You et al., 2009; Böhme et al., 2011). This climate event lasted for approximately 4 million years. Around 14 Ma Earth plunged into another cooling trend, and at 10 Ma the Arctic glaciations started. This cooling trend was almost continuous, with just a small warming interval during the mid Piacenzian, around 3 Ma (read Prista et al., 2014b). At 2.7 Ma the Northern Hemisphere continental glaciations begun, and Earth plunged into the present Ice Age (Naafs et al., 2010).

**Galactic environment and Cenozoic climate**
Earth processes haven't been able to completely explain Cenozoic climate evolution. However, if we look at the galactic environment throughout this Era, some coincidences are found. Optimum climates tend to occur outside the arms, while when crossing an arm region climate tend to globally cool. This would be expected from the cosmic rays link to the atmospheric current. Stronger CRF, stronger atmospheric current, increase in cloud formation. This sequence would lead to a global cooling. Looking at figure 1, this relation between the energy flux from Supernovae (SN) (main source of GCR) and Earth climate appears has something to be considered. Other studies suggest that this mechanism can't be ignored for long-term climate studies (geological time) (Shaviv, 2005; Medvedev and Melott, 2007; Kataoka *et al*., 2013).

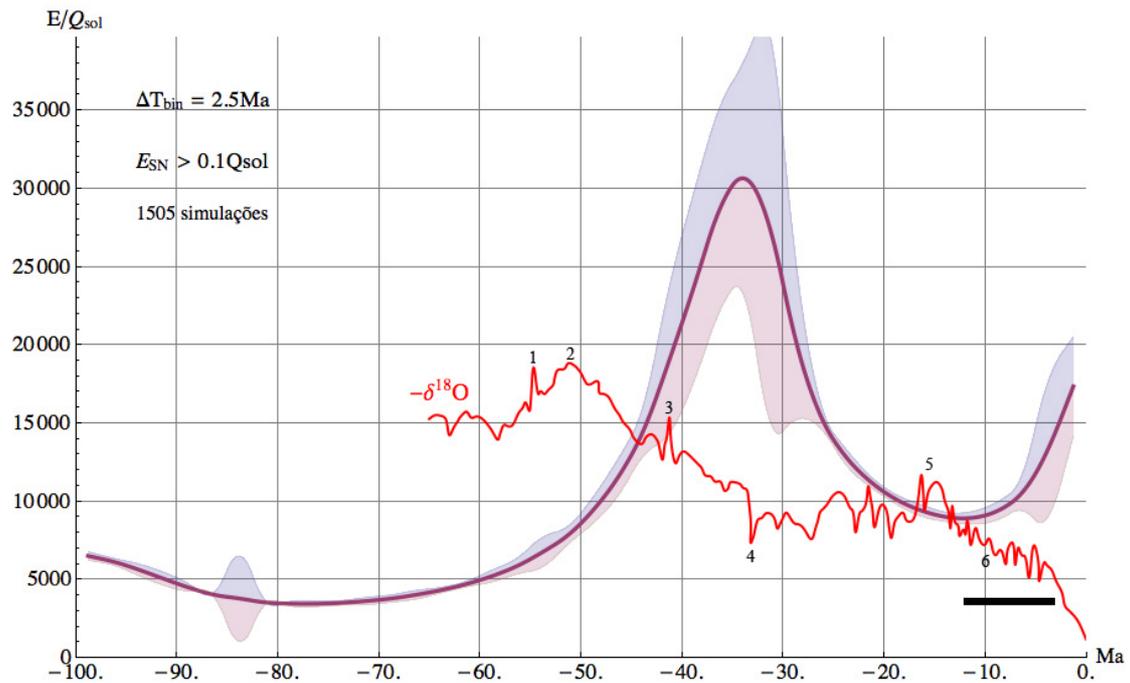

Figure 1 - Energy flux from the occurrences of SN in the Milky Way over the past 100 Ma. In red is the $\delta^{18}O$ curve from Zachos et al. (2008). 1- PETM; 2- EEOC; 3- MEOC; 4- ACG; 5- MCO; 6- Arctic glaciations; Black Bar- period in which 19 Supernovae occurred inside the Gould Region (Breitschwerdt and Avillez 2006).

**Conclusions**

Regarding the galactic environment: 1) optimum climates tend to occur outside the spiral arms (PETM, EEOC, MEOC and MCO); 2) the onset of the Antarctic glaciations is coincident with the passage through Sagittarius arm (the Antarctic Circumpolar Current was not formed yet and this suggests that the major increase in energy flux from SN may have played a role); 3) the beginning of ice formation in the Arctic region is coincident with the approximation to the Orion arm and the first occurrences of the 19 SN that occurred in the Gould region in the past 14 Ma; 4) climate degradation and the onset of the Northern Hemisphere continental glaciations is coincident with entering into the Orion arm and the occurrence of more SN in the Gould region.

Regarding Sirenia: its biodiversity was higher during the Eocene, got very low during the Oligocene, had a small peak during the MCO and dropped again with the subsequent global cooling (data from Prista el., 2014a).

Regarding climate: seasonality latitudinal gradient follows the energy flux variations. However it can been seen in figure 2 that other forces play a major role in this process, since seasonality is must higher with the lowest energy flux of the Orion Arm when compared to the highest flux of the Sagittarius Arm.

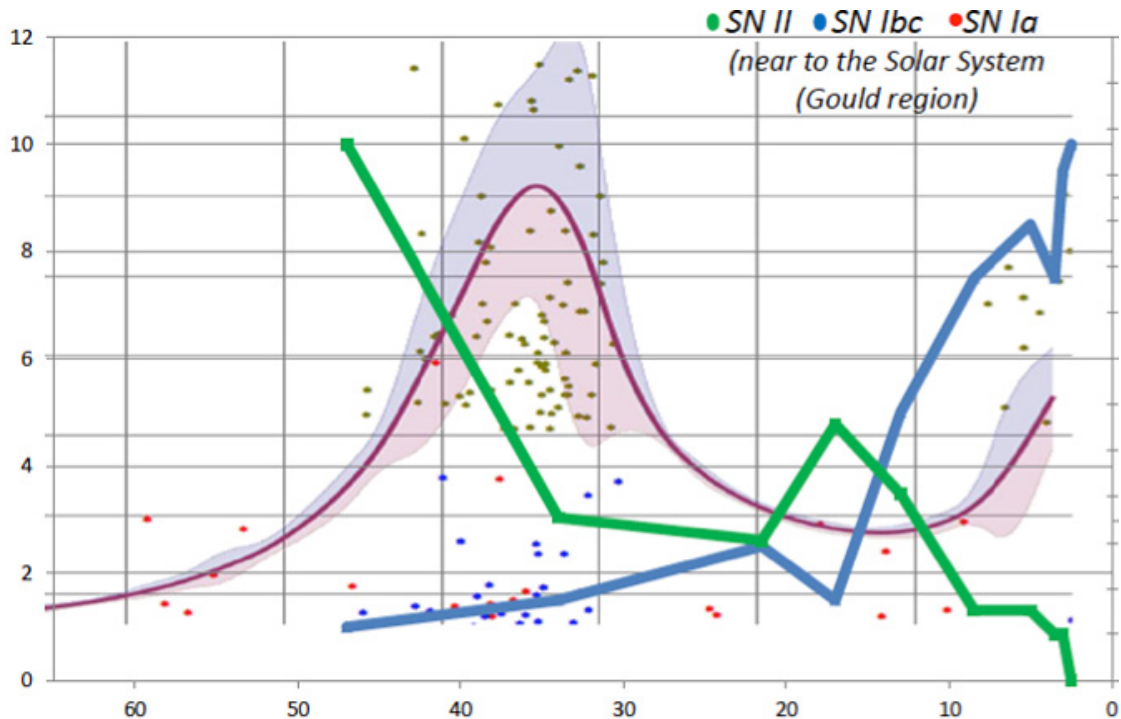

Figure 2 - Energy flux from SN in the Milky Way. Red dots - SNIa; Blue dots – SNIbc; Green dots – SNII. Green line – Sirenia biodiversity variation; Blue line – latitudinal temperature/seasonality variation.

Despite not being the only mechanism present, an influence of the galactic environment can be seen in long-term climate studies and should be taken into account. Indirectly, the Milky Way contributed to the disappearance of Sirenia from European shores.